\documentclass[a4paper,11pt]{article}

\usepackage{pos}
\usepackage[left]{lineno}
\usepackage{lipsum} 

\title{Search for Neutrinos from the Galactic 4FGL Sources with the Pion-bump Signature with IceCube}

\ShortTitle{IceCube: Search for Galactic Neutrino Sources}

\author{The IceCube Collaboration \\{\normalsize \normalfont(a complete list of authors can be found at the end of the proceedings)}\\}

\emailAdd{agranados@icecube.wisc.edu}
\emailAdd{rbabu@icecube.wisc.edu}
\emailAdd{mnisa@icecube.wisc.edu}

\abstract{

The IceCube Neutrino Observatory, located at the South Pole, covers a cubic kilometer of Antarctic ice, and is designed to detect astrophysical neutrinos in the TeV-PeV energy range. While IceCube has recently identified a diffuse flux of neutrinos originating from the Galactic Plane, specific sources of astrophysical neutrinos within the Milky Way remain elusive. Hadronic gamma-rays, produced through the decay of neutral pions, are expected to display a characteristic “pion bump” or “spectral break” around 200 MeV. Recent studies by the Fermi-LAT Collaboration highlight 56 sources from the 4FGL Catalog exhibiting a spectral break in the MeV energy range. Detecting astrophysical neutrinos from these sources would provide compelling evidence for cosmic-ray acceleration in their vicinity. In this analysis, we search for astrophysical neutrino emission from 56 sources showing characteristics of a pion bump using 13 years of IceCube data. Our findings could enhance our understanding of potential cosmic-ray acceleration sites in the galaxy.

\vspace{4mm}

{\bfseries Corresponding authors:}
 
Alejandra Granados$^{1*}$, 
Rishi Babu$^{1}$, 
Mehr Un Nisa$^{1}$\\

{$^{1}$ \itshape Dept. of Physics and Astronomy, Michigan State University, East Lansing, MI 48824, USA}\\[4mm]
$^*$ Presenter
}

\ConferenceLogo{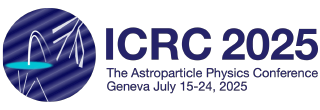}

\FullConference{39th International Cosmic Ray Conference (ICRC2025)\\
 15–24 July 2025\\
Geneva, Switzerland\\}

\begin{document}

\maketitle

\section{Introduction}\label{sec1}

The IceCube Neutrino Observatory, a cubic-kilometer detector located at the South Pole, has been searching for astrophysical neutrinos since 2010 \cite{Aartsen:2016nxy}. In 2023, IceCube detected diffuse neutrino emission from the Galactic Plane (GP) \cite{Aalbert:2023nxy}, confirming that high-energy neutrinos are being produced within our galaxy. However, no individual point sources have been identified to date. Detecting a Galactic neutrino point source would help address one of the longest-standing questions in high-energy astrophysics: the origin of cosmic rays.

The spectra of hadronic gamma rays, resulting from pion decay, exhibit a characteristic "pion bump" or spectral break near 200 MeV \cite{Abdollahi:2022yy}. High-energy protons, such as cosmic rays, interact with the interstellar medium, producing secondary particles such as pions. These pions subsequently decay into gamma rays and neutrinos. The presence of the pion bump serves as a distinct signature of hadronic emission. Sources displaying this feature are also potential neutrino emitters, as neutrinos are typically produced in hadronic interactions.

In this analysis, we search for astrophysical neutrino emission from 56 sources that exhibit characteristics of a pion bump, using 13 years of IceCube data. Our goal is to test if the spectral break indeed originates from hadronic processes, and to asses how much of the observed Galactic Plane flux could originate from these sources. To do this, we compare our sensitivities with the diffuse flux from the galactic plane reported in the recent IceCube analysis \cite{Aalbert:2023nxy}. These findings will help improve our understanding of potential cosmic-ray acceleration sites in the Galaxy.

\section{Potential Sources in the Galactic Plane}\label{sec2}

The Fermi Large Area Telescope (LAT) has identified 56 sources in the Galactic Plane from the 4FGL \cite{fermilat2020} that exhibit the characteristic "pion bump" spectral break between 50 MeV and 1 GeV \cite{Abdollahi:2022yy}. The source list includes the following classes: Supernova Remnants (SNR), Pulsar Wind Nebulae (PWN), High-Mass X-Ray Binaries (HMB), Star-Forming Regions (SFR), Supernova Remnants/Pulsar Wind Nebulae/composite sources (SPP), Binaries (BIN), Unidentified (UNID), and Unknown (UNK), shown in Table~\ref{tab:source_classes}. 

\begin{table}[h!]
\centering
\begin{tabular}{|l|c|}
\hline
\textbf{Source Class} & \textbf{\# of Sources} \\
\hline
Supernova Remnant (SNR) & 13 \\
High Mass X-Ray Binaries (HMB) & 3 \\
Pulsar Wind Nebulae (PWN) & 2 \\
Star Forming Regions (SFR) & 1 \\
Supernova Remnant/ Pulsar Wind Nebulae (SPP) & 6 \\
Binaries (BIN) & 1 \\
Unidentified (UNID) & 26 \\
Unknown (UNK) & 4 \\
\hline
\end{tabular}
\caption{Classification of the 56 sources used in this analysis.}
\label{tab:source_classes}
\end{table}

SPPs are sources of unknown nature but overlap with known SNRs or PWNe, making them candidates for these classes, while UNK are sources associated with counterparts of unknown nature \cite{Abdollahi:2022yy}. The UNID source class, on the other hand, contains sources that have not been definitely associated with any known astrophysical object or source type. Many of these classes, such as SNRs and PWNe, are known or suspected sites of cosmic acceleration \cite{Abdollahi:2022yy}, making them promising candidates for hadronic neutrino emission. Notably, a large fraction of the sources in the catalog remain unidentified, and this analysis contributes to their characterization regardless of whether a hadronic interpretation is confirmed.

\begin{figure}[t!]
\centering
\includegraphics[width=0.65\linewidth]{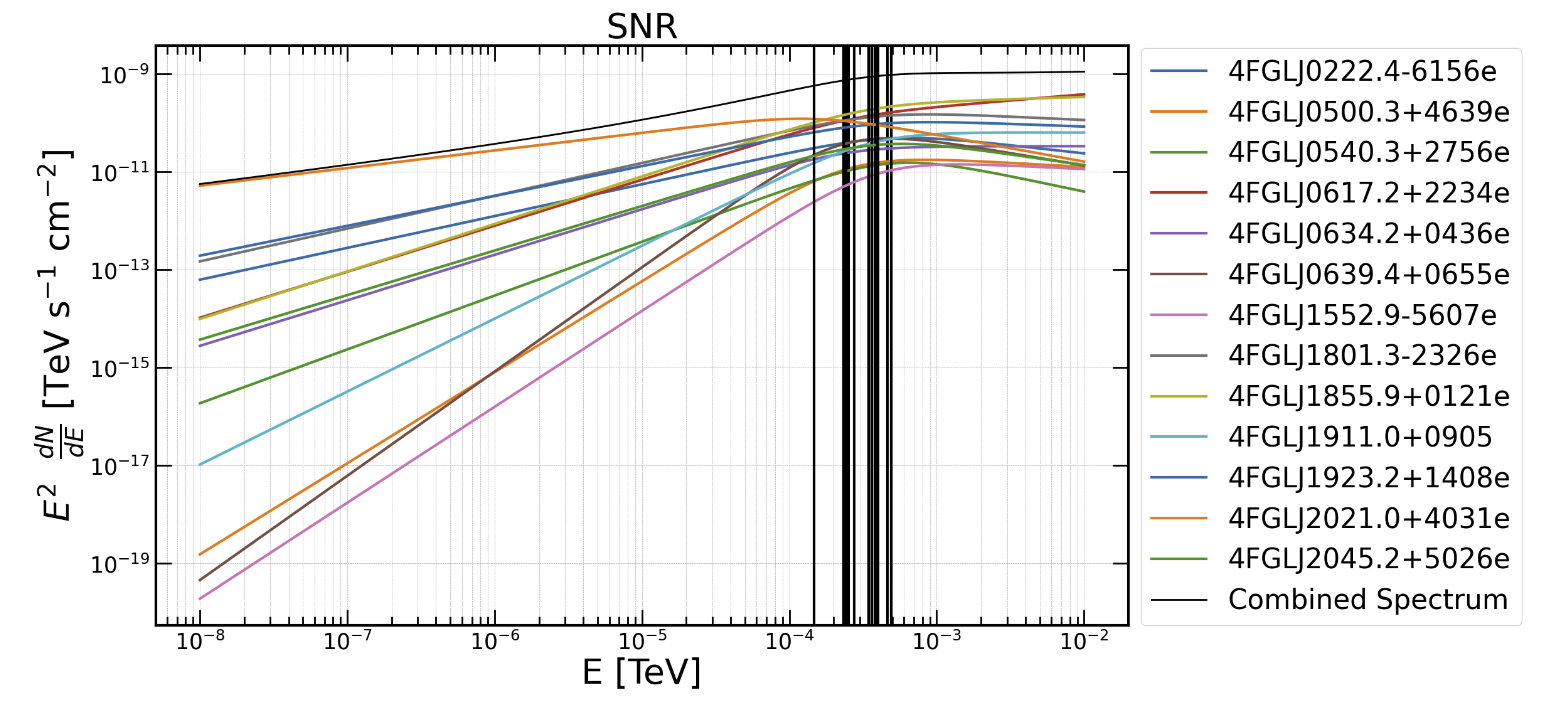}
\caption{\textbf{Gamma-ray spectral break for the SNR source class.} The spectral break resembles a pion-bump. The black lines represent the spectral break energies.}\label{fig:snr-fermi-spectrum}
\end{figure}

In addition, we cross referenced Fermi-LAT sources with the TeV Gamma-ray Catalog (TeVCat) \cite{2008ICRC....3.1341W} by searching for counterparts within $0.5^\circ$ of the Fermi source location. Of the 56 sources, 15 have a TeV counterpart. Since IceCube's sensitivity is greatest in the TeV range, these sources are especially relevant for neutrino searches. Table~\ref{tab:tev-counterparts} lists the 15 sources with TeV gamma-ray counterparts.

\begin{table}[ht]
\centering
\caption{List of 4FGL sources with TeV counterparts found within $0.5^\circ$.}
\begin{tabular}{|l|l|l|}
\hline
\textbf{4FGL Name} & \textbf{Source Class} & \textbf{TeVCat Counterpart} \\
\hline
4FGLJ0617.2+2234e & SNR & IC 443 \\
4FGLJ1801.3-2326e & SNR & W28 \\
4FGLJ1911.0+0905f & SNR & W49B \\
4FGLJ1923.2+1408e & SNR & W51 \\
4FGLJ1633.0-4746e & SPP & HESS J1632-478 \\
4FGLJ0340.4+5302 & UNID & LHAASO J0341+5258 \\
4FGLJ2108.0+5155 & UNID & LHAASO J2108+5153u \\
4FGLJ1018.9-5856 & PWN & HESS J1018-589B \\
4FGLJ1514.2-5909e & PWN & MSH15-52 \\
4FGLJ1857.7+0246e & PWN & HESS J1857+026 \\
4FGLJ0240.5+6113 & HMXB & LSI+61303 \\
4FGLJ2032.4+4053 & HMXB & LHAASO J2031+4052u \\
4FGLJ0545.1-5139 & BIN & Eta-Carinae \\
4FGLJ2028.6+4110e & SFR & Cocoon \\
4FGLJ1839.4-0553 & UNK & LHAASO J1839-0545 \\
\hline
\end{tabular}
\label{tab:tev-counterparts}
\end{table}

\section{Analysis Methods}\label{sec3}

\subsection{Detecting Neutrinos with IceCube} 

IceCube detects neutrinos by observing the Cherenkov light produced when secondary particles travel through the ice. The energy and direction of the incoming neutrinos are reconstructed based on the timing and intensity of light detected by the optical sensors \cite{Amanda2008}, \cite{IceCube2008}.

There are two primary types of neutrino events in IceCube. The first, known as \textit{tracks}, originate from charged-current $\nu_\mu$ interactions, which produce high-energy muons that travel long distances through the detector. These muons leave behind elongated Cherenkov light patterns, enabling precise angular reconstruction with a resolution of about $1^\circ$, making them ideal for point-source searches \cite{PSTracks2020}. The second type, called \textit{cascades}, result from $\nu_e$, $\nu_\tau$, and neutral-current interactions. These events produce roughly spherical light patterns from particle showers. While cascades offer excellent energy resolution, their angular resolution is coarser, approximately $10^\circ$, making them better suited for diffuse flux measurements \cite{DNNCascades2021}.

\begin{figure}[t!]
\centering
\includegraphics[width=0.6\linewidth]{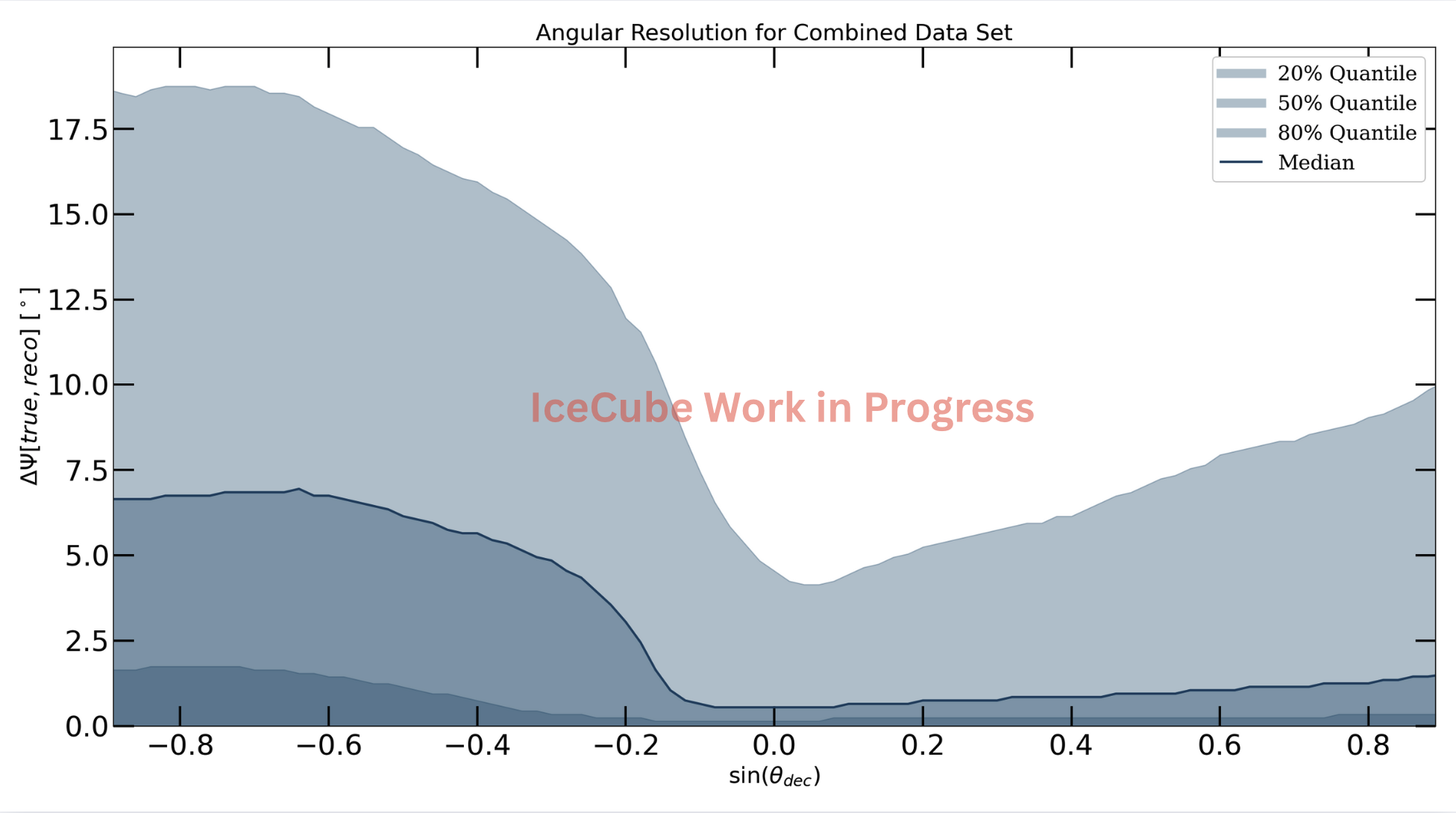}
\caption{\textbf{Angular resolution as a function of sin($\delta$) for the combined data set.} The plot shows the $20\%$ (darkest band), $50\%$ (middle band), and $80\%$ (lightest band) quantile which represents the containment of the angular separation between the true and reconstructed directions.}\label{fig:ang-res}
\end{figure}

This analysis incorporates both track and cascade events from 13 years of IceCube data, enabling a combined search. To avoid double counting, overlapping events between the two datasets were removed. Figure~\ref{fig:ang-res} shows the angular resolution of the combined data sample as a function of declination. In the Southern sky, the signal is dominated by cascades, while in the Northern sky, it is dominated by tracks. The median angular resolution remains below $7.5^\circ$ in the Southern sky and below $1^\circ$ in the Northern sky, reflecting the superior angular precision of track-like events. 

\subsection{Likelihood Method}

This analysis is based on the standard unbinned likelihood framework described in \cite{JBraun2008}, which has been widely used in IceCube point-source searches. The likelihood function depends on key parameters such as the number of signal neutrinos, the spectral index, and event observables such as the reconstructed energy and direction.

Within this framework, we perform two main types of analysis: a \textit{stacking analysis} and a \textit{catalog analysis}, as described in the following sections. To evaluate the presence of a signal, we define a test statistic (TS) that compares the likelihood of the signal-plus-background hypothesis to the background-only hypothesis. To estimate the background distribution, we randomize the data in right ascension (RA). Further, we mask the galactic plane when performing the randomization. This data-driven technique minimizes systematic errors while removing any real astrophysical signal, ensuring an unbiased sensitivity estimate. For each test, we perform 10,000 simulations on the scrambled data and obtain a TS distribution. To obtain a sensitivity, we inject an increasing number of signal events from Monte Carlo and record the resulting TS. We define the sensitivity flux as the flux level that exceeds the median background TS in $90\%$ of the background-only simulation.

\subsubsection{Stacking Analysis}

The stacking analysis combines information from multiple sources to enhance sensitivity by summing their contributions in a single likelihood. In this work, we stack sources by class, limiting the subsets to classes with more than five sources: SNR, SPP, UNID, and sources with TeV counterpart (TeVCat). Note that TeVCat is not a distinct source class but a cross-matched subset with TeV detections, included here as a separate stacking test to investigate the impact of TeV emission on detectability.

\begin{figure}[t!]
\centering
\includegraphics[width=0.6\linewidth]{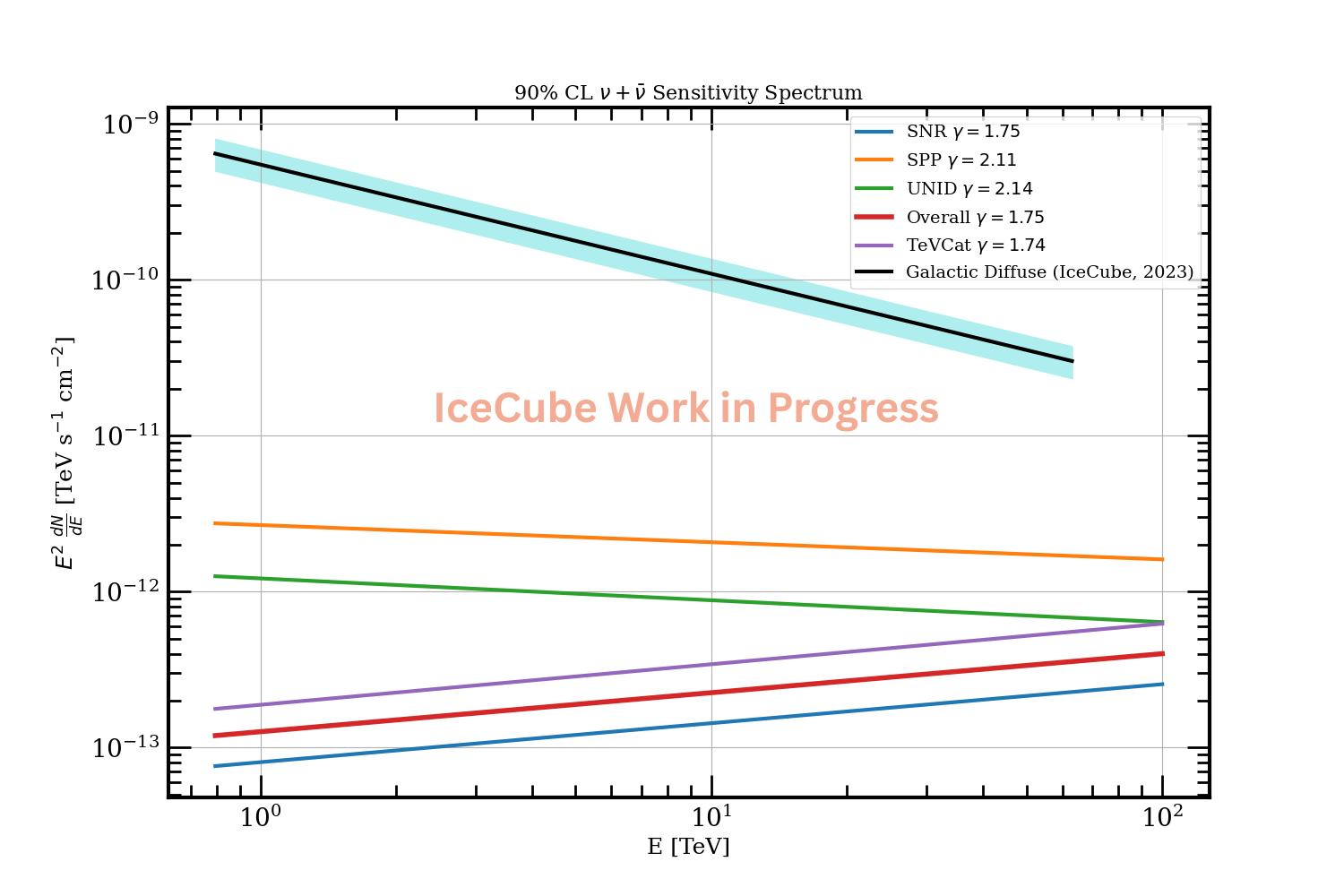}
\caption{\textbf{Stacked Sensitivity Spectra. The sources are weighted by their integrated ~MeV--GeV flux reported by Fermi.} For each of the source class, the integrated sensitivity is shown using the
$\gamma$ of the highest weighted source and $E_0 = 100$ TeV. The black solid line shows the $\pi^0$ best-fit flux from IceCube's Galactic plane diffuse measurement \cite{Aalbert:2023nxy} . The comparison with the Galactic plane diffuse emission illustrates the fraction of the Galactic diffuse flux that is constrained in the stacking analysis.}\label{fig:sens-spectrum}
\end{figure}

To compute the stacked sensitivity, we weight each source by the integrated MeV-GeV flux reported by  Fermi-LAT  \cite{Abdollahi:2022yy}. In addition, we incorporate the individual spectral indices ($\gamma$) of each source into the stacking when injecting a simulated flux. To report a representative sensitivity flux for each class, we adopt the spectral index of the most highly weighted source. Figure~\ref{fig:sens-spectrum} presents the sensitivity fluxes for each stacked class.

Our stacking results show sensitivities two orders of magnitude below the reported Galactic diffuse neutrino flux \cite{Aalbert:2023nxy}. This suggests that the analysis will be sensitive to a flux as weak as 1\% of the Galactic plane emission.

\subsubsection{Catalog Analysis}

\begin{figure}[t!]
\centering
\includegraphics[width=0.6\linewidth]{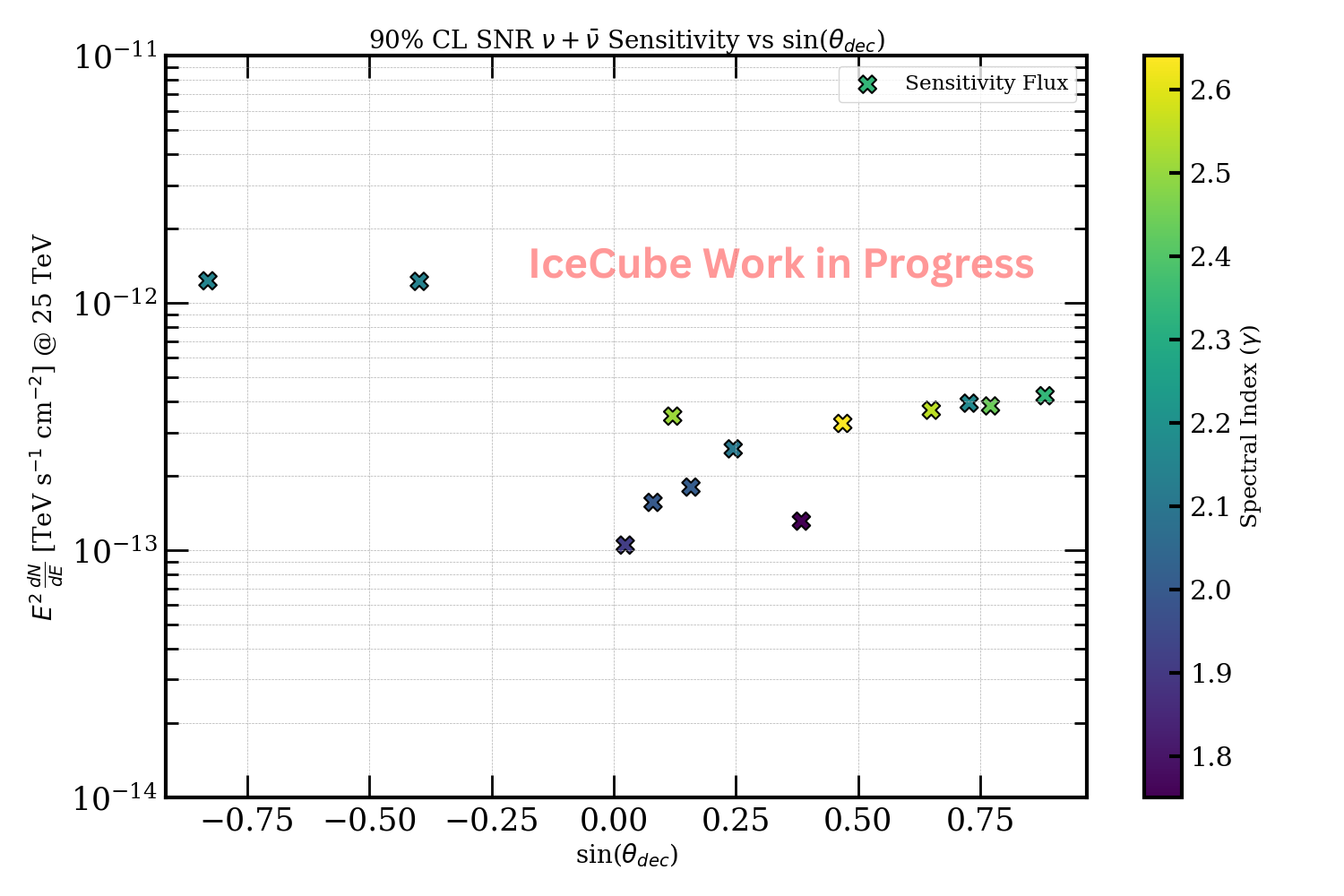}
\caption{\textbf{Sensitivity flux vs Sin ($\delta$) for sources in the SNR source class.} The 90\% confidence level of the sensitivity flux was calculated at 25 TeV, using the $\gamma$ from Fermi-LAT after the break.}\label{fig:snr-sens-catalog}
\end{figure}

The catalog analysis evaluates all 56 sources individually, and we compare their respective sensitivity fluxes to the measured diffuse flux. The sensitivity flux calculated for the SNR source class is shown in Figure~\ref{fig:snr-sens-catalog}.

\section{Outlook}\label{sec4}

This analysis provides a dedicated search for neutrinos from the Galactic Plane, focusing on sources having spectral features consistent with hadronic gamma-ray emission. By evaluating the neutrino contribution from 56 Fermi-LAT sources, we aim to constrain their contribution to the Galactic diffuse flux. Additionally, the results can contribute to the characterization of currently unidentified sources, regardless of whether a hadronic interpretation is ultimately confirmed.

\bibliographystyle{ICRC}
\bibliography{references}

\providecommand{\href}[2]{#2}\begingroup\raggedright\begin{thebibliography}{10}

\bibitem{Aartsen:2016nxy}
{\bfseries IceCube} Collaboration, M.~G. Aartsen {\em et~al.}, \href{http://dx.doi.org/10.1088/1748-0221/12/03/P03012}{{\em JINST} {\bfseries 12} no.~03, (2017) P03012}.

\bibitem{Aalbert:2023nxy}
{\bfseries IceCube} Collaboration, A.~Albert {\em et~al.}, \href{http://dx.doi.org/DOI:10.1126/science.adc9818}{{\em Science} {\bfseries 380} (2023) 1338--1343}.

\bibitem{Abdollahi:2022yy}
S.~Abdollahi {\em et~al.}, \href{http://dx.doi.org/10.3847/1538-4357/ac704f}{{\em ApJ} {\bfseries 933} (2022) 204}.

\bibitem{fermilat2020}
{\bfseries Fermi-LAT} Collaboration, S.~Abdollahi {\em et~al.}, \href{http://dx.doi.org/10.3847/1538-4365/ab6bcb}{{\em ApJS} {\bfseries 247} no.~33, (2020) }.

\bibitem{2008ICRC....3.1341W}
S.~P. {Wakely} and D.~{Horan}, {\em {International Cosmic Ray Conference}} {\bfseries 3} (2008) 1341--1344.

\bibitem{Amanda2008}
{\bfseries AMANDA} Collaboration, J.~Ahrens {\em et~al.}, \href{http://dx.doi.org/10.1016/j.nima.2004.01.065}{{\em Nucl.Instrum.Meth.A} {\bfseries 524} (2004) 169--194}.

\bibitem{IceCube2008}
{\bfseries IceCube} Collaboration, M.~Aartsen {\em et~al.}, \href{http://dx.doi.org/10.1088/1748-0221/9/03/P03009}{{\em JINST} {\bfseries 9} (2014) }.

\bibitem{PSTracks2020}
{\bfseries IceCube} Collaboration, M.~Aartsen {\em et~al.}, \href{http://dx.doi.org/10.1103/PhysRevLett.124.051103}{{\em Physical Review Letters} {\bfseries 124} (2020) }.

\bibitem{DNNCascades2021}
{\bfseries IceCube} Collaboration, R.~Abbasi {\em et~al.}, \href{http://dx.doi.org/10.1088/1748-0221/16/07/P07041}{{\em JINST} {\bfseries 16} (2021) P07041}.

\bibitem{JBraun2008}
J.~Braun {\em et~al.}, \href{http://dx.doi.org/10.1016/j.astropartphys.2008.02.007}{{\em Astroparticle physics} {\bfseries 29} (2008) 299--305}.

\end{thebibliography}\endgroup

\clearpage

\section*{Full Author List: IceCube Collaboration}

\scriptsize
\noindent
R. Abbasi$^{16}$,
M. Ackermann$^{63}$,
J. Adams$^{17}$,
S. K. Agarwalla$^{39,\: {\rm a}}$,
J. A. Aguilar$^{10}$,
M. Ahlers$^{21}$,
J.M. Alameddine$^{22}$,
S. Ali$^{35}$,
N. M. Amin$^{43}$,
K. Andeen$^{41}$
% ...
\\
\\
% ...
$^{10}$ Universit{\'e} Libre de Bruxelles, Science Faculty CP230, B-1050 Brussels, Belgium \\
% ...
$^{16}$ Department of Physics, Loyola University Chicago, Chicago, IL 60660, USA \\
$^{17}$ Dept. of Physics and Astronomy, University of Canterbury, Private Bag 4800, Christchurch, New Zealand \\
% ...
$^{21}$ Niels Bohr Institute, University of Copenhagen, DK-2100 Copenhagen, Denmark \\
$^{22}$ Dept. of Physics, TU Dortmund University, D-44221 Dortmund, Germany \\
% ...
$^{35}$ Dept. of Physics and Astronomy, University of Kansas, Lawrence, KS 66045, USA \\
% ...
$^{39}$ Dept. of Physics and Wisconsin IceCube Particle Astrophysics Center, University of Wisconsin{\textemdash}Madison, Madison, WI 53706, USA \\
% ...
$^{41}$ Department of Physics, Marquette University, Milwaukee, WI 53201, USA \\
% ...
$^{43}$ Bartol Research Institute and Dept. of Physics and Astronomy, University of Delaware, Newark, DE 19716, USA \\
% ...
$^{63}$ Deutsches Elektronen-Synchrotron DESY, Platanenallee 6, D-15738 Zeuthen, Germany \\
$^{\rm a}$ also at Institute of Physics, Sachivalaya Marg, Sainik School Post, Bhubaneswar 751005, India \\
% ...

\subsection*{Acknowledgments}

\noindent
The authors gratefully acknowledge the support from the following agencies and institutions:
USA {\textendash} U.S. National Science Foundation-Office of Polar Programs,
U.S. National Science Foundation-Physics Division,
U.S. National Science Foundation-EPSCoR,
U.S. National Science Foundation-Office of Advanced Cyberinfrastructure,
Wisconsin Alumni Research Foundation,
Center for High Throughput Computing (CHTC) at the University of Wisconsin{\textendash}Madison,
Open Science Grid (OSG),
Partnership to Advance Throughput Computing (PATh),
Advanced Cyberinfrastructure Coordination Ecosystem: Services {\&} Support (ACCESS),
Frontera and Ranch computing project at the Texas Advanced Computing Center,
U.S. Department of Energy-National Energy Research Scientific Computing Center,
Particle astrophysics research computing center at the University of Maryland,
Institute for Cyber-Enabled Research at Michigan State University,
Astroparticle physics computational facility at Marquette University,
NVIDIA Corporation,
and Google Cloud Platform;
Belgium {\textendash} Funds for Scientific Research (FRS-FNRS and FWO),
FWO Odysseus and Big Science programmes,
and Belgian Federal Science Policy Office (Belspo);
Germany {\textendash} Bundesministerium f{\"u}r Forschung, Technologie und Raumfahrt (BMFTR),
Deutsche Forschungsgemeinschaft (DFG),
Helmholtz Alliance for Astroparticle Physics (HAP),
Initiative and Networking Fund of the Helmholtz Association,
Deutsches Elektronen Synchrotron (DESY),
and High Performance Computing cluster of the RWTH Aachen;
Sweden {\textendash} Swedish Research Council,
Swedish Polar Research Secretariat,
Swedish National Infrastructure for Computing (SNIC),
and Knut and Alice Wallenberg Foundation;
European Union {\textendash} EGI Advanced Computing for research;
Australia {\textendash} Australian Research Council;
Canada {\textendash} Natural Sciences and Engineering Research Council of Canada,
Calcul Qu{\'e}bec, Compute Ontario, Canada Foundation for Innovation, WestGrid, and Digital Research Alliance of Canada;
Denmark {\textendash} Villum Fonden, Carlsberg Foundation, and European Commission;
New Zealand {\textendash} Marsden Fund;
Japan {\textendash} Japan Society for Promotion of Science (JSPS)
and Institute for Global Prominent Research (IGPR) of Chiba University;
Korea {\textendash} National Research Foundation of Korea (NRF);
Switzerland {\textendash} Swiss National Science Foundation (SNSF).

\end{document}